%% file: main.tex
\documentclass[times,12pt]{article}

\usepackage[letterpaper, portrait, margin=1in]{geometry}
\usepackage[utf8]{inputenc}
\usepackage{amsmath} 
\usepackage{amssymb}
\usepackage{graphicx}
\usepackage{setspace}
\usepackage{parskip}
\usepackage{array}
\usepackage{booktabs}
\usepackage{hanging}
\usepackage{natbib}
\usepackage{authblk}
\usepackage{lineno}

\title{Shear-enhanced permeability development of magma vesiculating in cylindrical conduits}
\author[a,*]{J. Birnbaum}
\author[b]{J. Schauroth}
\author[b]{J. Weaver}
\author[a]{J.E. Kendrick}
\author[a]{A. Lamur}
\author[a]{Y. Lavall\'ee}
\affil[a]{ Department of Earth and Environmental Sciences, Ludwig-Maximilians-Universit\"at M\"unchen, Theresienstra{\ss}e 41, 80333 Munich, Germany}
\affil[b]{ Earth, Ocean and Ecological Sciences, University of Liverpool, 4 Brownlow Street, Liverpool, United Kingdom L69 3GP}
\affil[*]{Corresponding Author: janine.birnbaum@min.uni-muenchen.de}

\begin{document}

%TC:ignore

\maketitle
\begin{abstract}
    During magma vesiculation, permeability is established when growing bubbles begin to form connected networks, which allow fluids to percolate. This percolation threshold controls the relative rates between magma ascent and volatile exsolution, which in turn dictate eruptive style. Percolation is controlled primarily by total vesicularity and shear conditions. We performed vesiculation experiments on rhyolitic glass in a spatially confined, cylindrical, conduit-like geometry. The amount of shear experienced by the sample is controlled by varying the sample and confining diameters to allow for various degrees of free (isotropic) followed by confined (anisotropic) expansion. Pore anisotropy develops sub-parallel to the flow direction. We measure the total and connected porosity and permeability of the vesiculated samples. We observe two regimes of behavior for samples dominated by 1) isotropic expansion, in which the onset of percolation corresponds with the beginning of shear deformation and 2) anisotropic expansion which shows low percolation thresholds ($<$20\%) with a near-constant permeability even with increasing porosity. We find a good correlation between connected porosity and permeability, with distinct trends for sheared and unsheared samples. The development of anisotropy in vesiculating magmas is ubiquitous in magmatic networks and our data highlight the importance of considering the in situ shear conditions when investigating the percolation threshold.

    \vspace{3 mm}
    \noindent \textbf{Keywords:} Magma vesiculation,  Bubble shear, Porosity, Permeability
\end{abstract}

\section*{Highlights}
\begin{itemize}
    \item Shear in vesicular magma promotes pore connectivity which persists after shearing ends.
    \item Permeability is higher in the presence of shear during vesiculation.
    \item After percolation onset, permeability is stable even when porosity increases.
\end{itemize}

%TC:endignore

\section{Introduction}
The transport and eruption of magma is driven by buoyancy, imparted primarily by volatile exsolution. Examining the eruptive record, eruptions of all sizes can shift between effusive and explosive modes, or exhibit a combination of the two \citep[e.g.][]{Matthews1997,Schipper2013}, depending on the ability and efficiency of magma to release gas pressure \citep{Eichelberger1986, Jaupart1991, Woods1994, Degruyter2012, Gonnermann2013}, controlled by its vesicularity and, importantly, permeability. It is therefore crucial to understand the processes and timescales of volatile loss from magma via vesiculation (the combination of bubble nucleation, growth, and coalescence) and outgassing. \par

The solubility of volatiles (e.g., H$_2$O, CO$_2$) in magmas is dependent on magma chemistry, pressure, and temperature. During eruptions, magmas rapidly decompress, reducing the solubility of volatile species prompting exsolution \citep{Sparks1978}. Shear-induced heating may also be a mechanism for decreasing the solubility of volatiles during eruptions \citep{Lavallee2015}. When magmas are oversaturated, volatiles exsolve initially via bubble nucleation, which requires relatively high degrees of oversaturation and is closely tied to high decompression rates or slow diffusion \citep{Toramaru2006} and the presence of crystals to act as nucleation sites \citep{Hurwitz1994, Shea2017}. If diffusion is relatively efficient, exsolution is primarily accommodated by the growth of existing bubbles. In comparison to bubble nucleation, bubble growth is better understood and numerous works have been dedicated to this process, including detailed analytical and numerical models \citep[e.g.][]{Prousevitch1993, Kaminski1997, Lensky2001, Ryan2015, Coumans2020b}. As bubbles grow, the melt films separating them may thin and eventually rupture, resulting in bubble coalescence. Although coalescence is a process of considerable interest, the physical processes of bubble coalescence are complex \citep[e.g.][]{Castro2012a}, and crystallinity \citep{Oppenheimer2015} and shear conditions \citep{Okumura2006} likely play important roles. \par

As bubbles grow closer and interact or coalesce, it becomes possible for gas to flow between bubbles or  diffuse through sufficiently thin melt films \citep[e.g.][]{Rust2004,Mueller2005}. This connectivity threshold for percolation allows for permeable outgassing and subsequent densification of the magma in the ``permeable foam'' model of silicic magma transport \citep{Eichelberger1986, Fink1992, Westrich1994, Tuffen2003}. An idealized magma comprised of randomly placed fully penetrable spherical bubbles predicts a critical porosity of $\sim$30\% \citep{Blower2001,Vasseur2017}. Laboratory vesiculation experiments have found a higher threshold for connectivity attributed to the non-random locations of bubbles during nucleation and growth, or viscous barriers to coalescence \citep{Takeuchi2008, Takeuchi2009}. If instead we assume bubbles can be approximated as ordered, non-overlapping rigid spheres, maximum packing is $\approx64\%$, for an upper bound. Furthermore, uncoalesced bubbles in close proximity may take on polygonal shapes, which can support very high volume fractions of gas which are observed in some volcanic eruption products (e.g., reticulite).  \par

In reality magmas generally contain variable fractions of bubbles and crystals with complex shape and size distributions, and their properties are scale-dependent and rarely homogeneous or isotropic. Measurements of natural and experimental samples show percolation thresholds between $\sim$30-80\% \citep[e.g.][]{Burgisser2017}. The percolation threshold and permeability of volcanic rocks vary as a result of shear deformation \citep{Burgisser2004, Takeuchi2005, Okumura2006, BouvetdeMaisonneuve2008, Okumura2008, Schipper2013, Gonnermann2017}, strain localization and fracture development \citep{Okumura2009, Okumura2010, Laumonier2011, Kendrick2013, Lavallee2013, Ashwell2015, Kushnir2017, Lamur2017},  melt chemistry \citep{Lindoo2016}, magma rheology \citep{Castro2012a, Colombier2020}, the presence of crystals \citep{Bai2011, Lindoo2017}, the distribution of bubble and crystal microstructure \citep{Saar1999, Burgisser2017, Colombier2017, Giachetti2019, Colombier2020}, and time \citep{Baker2012}. \par

Beyond the percolation threshold, permeability shares non-linear relationships with porosity (which may consists of vesicles, intergranular voids, and cracks). In natural materials, large variations in permeability exist for a given porosity, suggesting that the construction of porous networks in magma is complex, due to varying eruptive histories \citep{Klug1996, Mueller2005, Wright2009a, Nguyen2014, Farquharson2015, Colombier2017}.  Effusive and explosive eruptive products exhibit broadly different permeability-porosity relationships, underlined by differing dominant porosity-controlling mechanisms with a hysteresis effect \citep{Rust2004, Lindoo2016, Gonnermann2017}; explosive products display permeable pathways apparently controlled by bubble growth and coalescence, which can be modeled by percolation theory, whereas effusive products have permeable pathways controlled by shear, bubble elongation, compaction, fracturing, and welding, are better approximated by a combination of capillary tube \cite[e.g., via the Kozeny-Carman relationship;][]{Saar1999, Costa2006} and fracture models \citep{Mueller2005}. These models have further been adapted, realizing that magma may fragment and sinter repeatedly during ascent \citep{Wadsworth2020}, thus the porous permeable network reflects a combination of these processes operating at different times and locations. \par 

In natural systems, the processes of porosity creation and destruction will vary in space and time, with profound impacts on eruption rate and style. This will result in variable vesicle textures, reflecting e. g. the distribution of shear imparted on vesicular magmas during emplacement, even within a single eruptive unit \citep{BouvetdeMaisonneuve2008}. Understanding the complex evolution of porosity and permeability in response to vesiculation, shear, and outgassing remains an outstanding challenge. Despite its relevance to volcanological settings, the development of permeability in conduit-like geometries remains understudied. Our experiments allow for simultaneous vesiculation, permeable flow, and shear characterized by a combination of initially isotropic expansion, followed by anisotropic expansion that results in different bubble shape distributions and porosity evolution. \par

\section{Materials and methods}
\subsection{Experimental setup}
The development of permeability in vesiculating magma was studied here using two test methods 1) unconfined vesiculation promoting isotropic expansion, and 2) variably confined vesiculation promoting momentary isotropic expansion followed by different extents of anisotropic expansion. For both test types we prepared and used cylindrical cores of aphyric rhyolitic obsidian from Hrafntinnuhryggur ridge at Krafla volcano (Iceland), with diameters and heights (1:1 ratio) of 10 mm and 12 mm. For confined experiments, samples were enclosed in cylindrical shells (i.e., hollow tubes) of dense, holocrystalline basalt from Seljavellir (Iceland), as the material is stable at the experimental conditions and has a low permeability. \par

\subsection{Characterisation of obsidian samples}
Whole-rock geochemistry of the obsidian was determined by X-ray fluorescence (XRF) in a PANalytical Axios Advanced XRF spectrometer at the University of Leicester. Major elements were measured on glass beads fused from ignited powders using a sample to flux ratio of 1:5 (80\% Li metaborate: 20\% Li tetraborate). Results are reported as component oxide weight percent and have been recalculated to include loss on ignition (LOI). \par

The water concentration dissolved in the rhyolitic glass was measured by Fourier-transform infrared spectroscopy (FTIR) using a Thermo Nicolet 380 FT-IR and Nicolet Centaurus microscope at the University of Liverpool. Doubly-polished wafers were prepared with a thickness of $\sim$130 $\mu$m, determined with a high- precision calliper. The spectra of four spots, $\sim$100 × 100 $\mu$m, were acquired and analysed. H$_2$O$_t$ concentrations were calculated by measuring the absorbance of the $\sim$3500 cm$^{-1}$ peak above a linear background and using the Beer-Lambert law, with a molar absorptivity coefficient of 90 mol$^{-1}$cm$^{-1}$ \citep{Hauri2002} and a dry powder density of 2400 kg m$^{-3}$, measured by helium pycnometry. \par 

The glass transition temperature, T$_g$, was determined by differential scanning calorimetry (DSC) in a Netzsch simultaneous thermal analyser (STA) Jupiter 449 F1 at the University of Liverpool, using a 1 mm thickness, 6 mm diameter disk of rhyolite weighing 67 mg. Each test was initially performed on an empty lidded platinum crucible to provide a correction baseline, followed by measurement on the sample. The assembly was heated to 1000 $^\circ$C at 10 $^\circ$C min$^{-1}$ in the presence of a 20 mL min$^{-1}$ argon flow in the surrounding atmosphere. We identified T$_g$ as the temperature of the endothermic peak in the DSC curve. 
\par

\subsection{Unconfined vesiculation experiments}
\label{section:isotropic_methods}
To determine the development of porosity and permeability associated with vesiculation of a freely expanding sample, we heated obsidian cores (10 mm/ 12 mm) at 10 $^\circ$C min$^{-1}$ to 1009 $^\circ$C for different isothermal dwell times of 10–600 min in a Carbolite box furnace. The samples were cooled at a rate of 5 $^\circ$C min$^{-1}$, chosen to minimize the relaxation of the bubble textures, whilst being slow enough to avoid high thermal gradients across the cooling sample which could promote cracking. \par 

The volume of each sample was measured prior to ($V_\text{pre}$) and following ($V_\text{post}$) experimentation via helium pycnometry, to provide the total porosity: 
\begin{equation}
    \phi = 1 - \frac{V_\text{pre}}{V_\text{post}} \: . 
    \label{eq:vesicularity}
\end{equation}
We note that porosity includes both vesicularity and the void space occupied by fractures that may form both during shear in the experiments and also during cooling-induced contraction. We make the simplifying assumption that the porosity and permeability of our samples is dominated by the vesiculation and that the cooled samples are representative of the microstructure present during the experiment. Additionally, we assume that none of the porosity is connected to the exterior of the sample due to the presence of a dense and impermeable glassy rind formed by diffusive gas loss in a volumetrically insignificant margin around the sample \citep{vonAulock2017, Weaver2022}. Following determination of porosity, the samples were prepared for permeability measurements (see Section \ref{section:permeability_methods}). \par 

For a subset of experiments, the samples were imaged through a sapphire window in the furnace using a FLIR thermographic camera (sampling rate of 1 frame min$^{-1}$, spatial resolution of 31.9$\pm$4.2 pixels$\cdot$mm$^{-2}$) to monitor sample area (as a proxy for vesicularity) through time. From the images we identified the sample edges, and owing to the cylindrical geometry, we employed the solid of revolution method to geometrically convert the silhouettes into three-dimensional volumes. This approach assumes the sample remains vertically axisymmetric throughout the length of the experiment. \par 

%TC:ignore
\begin{figure}[]
\begin{center}
\includegraphics[width=\textwidth]{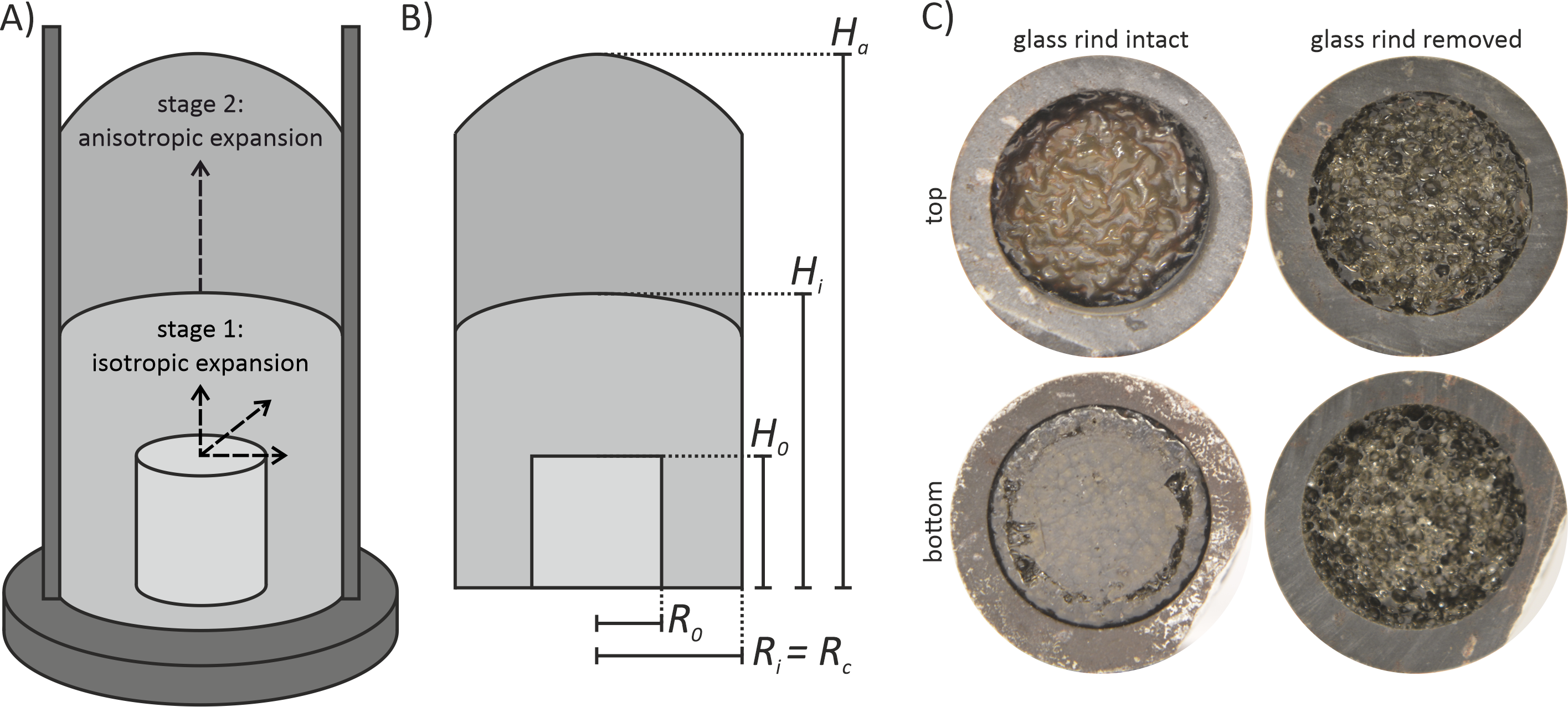}\\
\caption{A) Sketch of experimental setup for confined experiments; B) definitions of radius and height for the initial sample ($R_0$ and $H_0$, respectively), the maximum extent of isotropic expansion ($R_i$ and $H_i$), and the anisotropic expansion ($R_a$ and $H_a$); C) photographs of post-experimental samples from above and below, before and after removing the glass rind.}
\label{fig:setup}
\end{center}
\end{figure}
%TC:endignore

\subsection{Confined vesiculation experiments}
\label{section:anisotropic_methods}

To determine the impact of spatial confinement on vesiculating magma in a conduit-like geometry, samples were confined in open-topped cylindrical crucibles of dense, holocrystalline basalt. The crucibles all have an external diameter of 25 mm, with various inner diameters  of 10.5, 12.7, 14.5, and 18.6 mm. By combining different sample radii (5 or 6 mm), $R_0$, and inner tube radii, $R_c$, we achieved a range of variably confined axisymmetric geometries (see Fig. \ref{fig:setup}A\&B for sketches of the experimental setup). We define the radial confinement ratio, C:
\begin{equation}
    C = \frac{R_0}{R_c} \: , 
\end{equation}
to quantify how much the sample is able to expand isotropically before its margins touch and are impeded by the crucible; any further vesiculation forces expansion to occur along the long axis of the conduit, thus causing anisotropy. When $C$=1, vesiculation is fully anisotropic, and when $C<$1, vesiculation is first isotropic, then anisotropic. \par 

The samples were placed on a $\sim$1 cm thick, disk-shaped base plate of the same basalt. Prior to the vesiculation experiments, base plates and tubes were thermally treated at 1009 $^\circ$C for 8 h to avoid reaction of the basalt during the experiments. We cemented the basaltic tubes onto the base plates prior to the experiments using a thin film of high-temperature cement to prevent rupture of the assembly during sample vesiculation. \par 

Confined vesiculation experiments followed the same temperature path as the isotropic experiments (heating rate: 10 $^\circ$C min$^{-1}$, dwell temperature: 1009 $^\circ$C, cooling rate: 5 $^\circ$C min$^{-1}$). We varied the dwell time between 10-480 min. The volumes of the samples before and after the experiment were measured along with the basalt crucibles using helium pycnometry as above. \par

\subsection{Permeability measurements}
\label{section:permeability_methods}
After the porosity determination, samples fused to the crucible were cut at the top and bottom to remove the impermeable rind that forms during diffusive outgassing (Fig. \ref{fig:setup}C). Unconfined samples and samples not fully coupled to the crucible were cut and measured in a rubber jacket. Permeability through the cut samples was measured using a Vinci Technologies gas permeameter with nitrogen gas, inserted into a compressible Viton jacket and radially loaded to 1 MPa confining pressure using a manual valve. A steady gas flow rate through the sample was selected (2 to 125 cm$^3$ min$^{-1}$) to maintain a pressure differential across the sample of $\approx$0.5 psi. The sample permeability was calculated using Darcy’s law with a fluid viscosity for nitrogen at ambient temperature (1.741-1.780 $\times 10^{-5}$ Pa$\cdot$s), and the cross-sectional surface area of the vesiculated sample (from either the average sample diameter or the inner diameter of the cylindrical crucible for fused samples). For very low permeability samples, measurements terminated while the pressure gradient was still increasing, but within the machine tolerance for error and so represent maximum values that depend primarily on the sample dimensions. The permeability of the basalt crucibles is insignificant compared to the sample permeability; the obsidian has a permeability of 5$\times 10^{-20}$ m$^2$ \citep{Lamur2018}. After permeability measurements, the cut cores were again measured in the helium pycnometer to determine the connected porosity of the central portion of the sample using the computed cylindrical sample and crucible volumes. 

\subsection{Textural analysis}
\label{section:textural_methods}
We selected two confined and four unconfined samples for backscatter electron (BSE) imaging using a Hitachi TM3000 scanning electron microscope (SEM) at the University of Liverpool. The samples were cut along the vertical axis (parallel to the cylindrical conduit for the anisotropic experiments) using a Well 3500 precision diamond saw. Photos were merged using Adobe Photoshop®.

\section{Results}
\subsection{Obsidian characteristics}
The major element composition obtained via XRF analysis revealed a peraluminous rhyolitic glass and total water from FTIR analysis is 0.12 wt.\% (Table \ref{tab:composition}). DSC measurements had a peak associated with T$_g$ at 684 $^\circ$C, suggesting the experimental temperature of 1009 $^\circ$C ensures a relaxed liquid at the timescale of the experiments. Using the viscosity model of \citet{Hess1996} for hydrous leuco-granitic melts, we estimate a melt viscosity of 2.25$\times 10^7$ Pa$\cdot$s at the experimental dwell temperature. Using the low-pressure water solubility model of \citet{Ryan2015} for a hydrous rhyolitic melt, we estimated that 0.10 wt.\% H$_2$O is soluble in the melt at experimental conditions. \par

\input{Table_composition.tex}

\subsection{Unconfined vesiculation experiments}
\begin{figure}[]
\begin{center}
\includegraphics[width=5in]{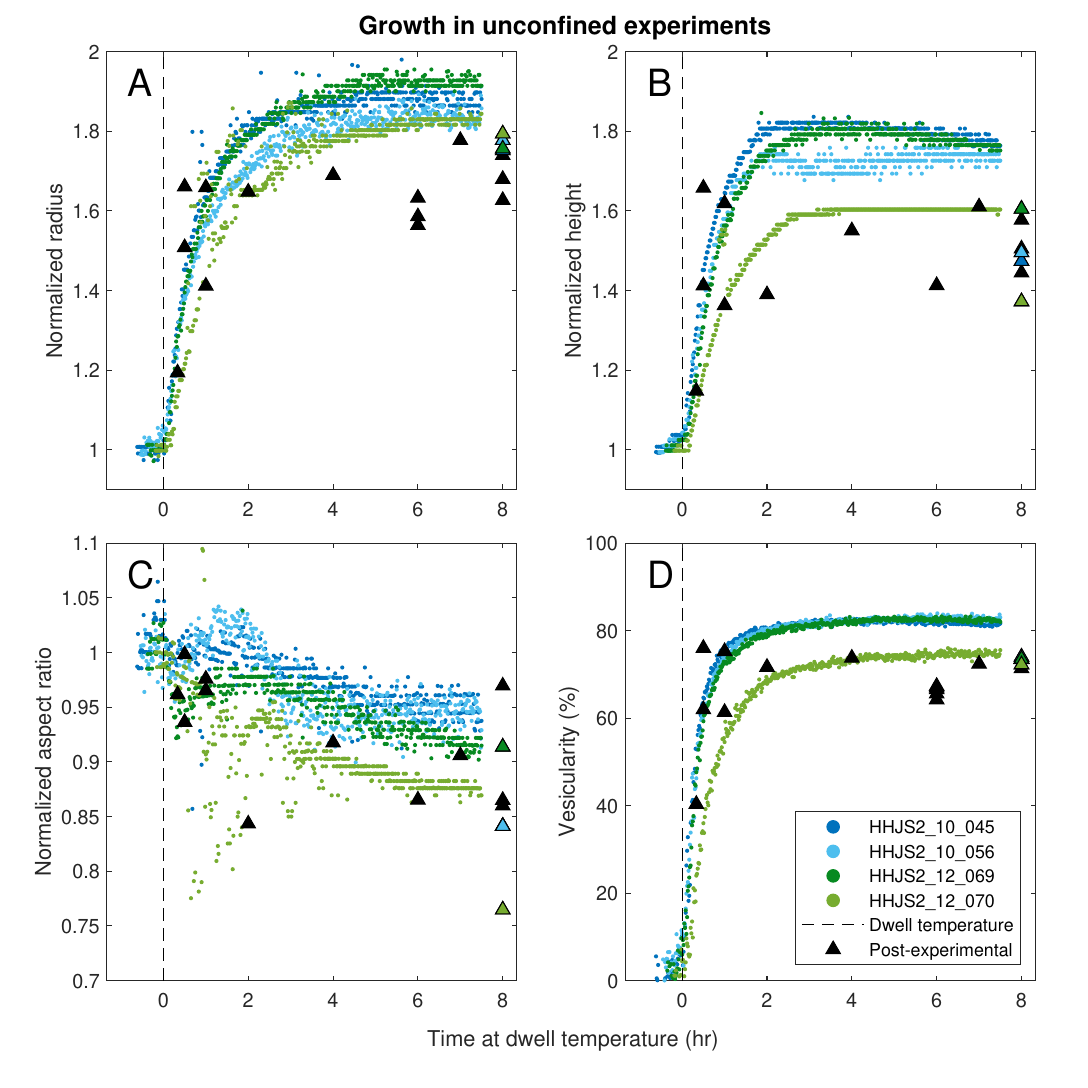}\\
\caption{Sample dimensions measured from FLIR imaging of unconfined experiments showing changes in A) normalized radius, B) normalized height, C) normalized aspect ratio, and D) vesicularity. Small markers show each frame colored by experiment number. Samples HHJS2\_10\_045 and HHJS2\_10\_056 had an initial radius of 10 mm and HHJS2\_12\_069 and HHJS2\_12\_070 had an initial radius of 12 mm, and all were unconfined. Triangles show post-experimental analyses colored for the same experiments and black for experiments not imaged with FLIR. Dashed line in all panels indicates the time at which experiments reach the dwell temperature of 1009 $^\circ$C.}
\label{fig:isotropic_results}
\end{center}
\end{figure}

We vesiculated samples in unconfined geometries to develop a baseline for how unsheared samples behave. We observe that vesiculation begins at a temperature of $\approx$950 $^\circ$C, 10 min before we reach the dwell temperature of 1009 $^\circ$C. By the time we reach the dwell temperature, the samples already contain 5-10\% vesicularity. This is accommodated by preferential growth in the vertical direction. At the dwell temperature, the vesiculation rate accelerates and then decreases, with most of the vesiculation occurring within 100 min at the dwell temperature, consistent with the findings and interpretations of \citet{Weaver2022}. Samples reach a maximum apparent vesicularity of 80-85 vol\%. The volumetric increase is accommodated by a near doubling of both apparent radius and height (Fig. \ref{fig:isotropic_results}). With increasing time, the samples continue to expand laterally at the expense of vertical growth, showing a gradual decrease in the aspect ratio. \par 

%TC:ignore
%TC:endignore

We find reasonable agreement in vesicularity measurements via syn-experimental thermal imaging and post-experimental analysis, although we may have lost a small amount of porosity due to bubble resorption and shrinkage on cooling (Fig. \ref{fig:isotropic_results}D). Discrepancy between the final radius ((Fig. \ref{fig:isotropic_results}A) and height ((Fig. \ref{fig:isotropic_results}B) may be partially attributed to thermal expansivity of the material, and differences in the measurement strategy; image analysis uses the 90th percentile of detected pixels while post-experimental analysis is a mean of 10 measurements across the sample (usually mostly near the middle and would have a stronger effect on height than radius). However, as our FLIR measurements only give reliable shapes until the end of the dwell period, we cannot rule out continued vesiculation and flow of the sample during the early parts of cooling, which may explain the systematic decrease in height and aspect ratio on the final samples and why the final size and porosity of samples with short dwell times are higher than the FLIR measurements after the same dwell time. \par

\subsection{Porosity and permeability of confined experiments}
\begin{figure}[]
\begin{center}
\includegraphics[width=4in]{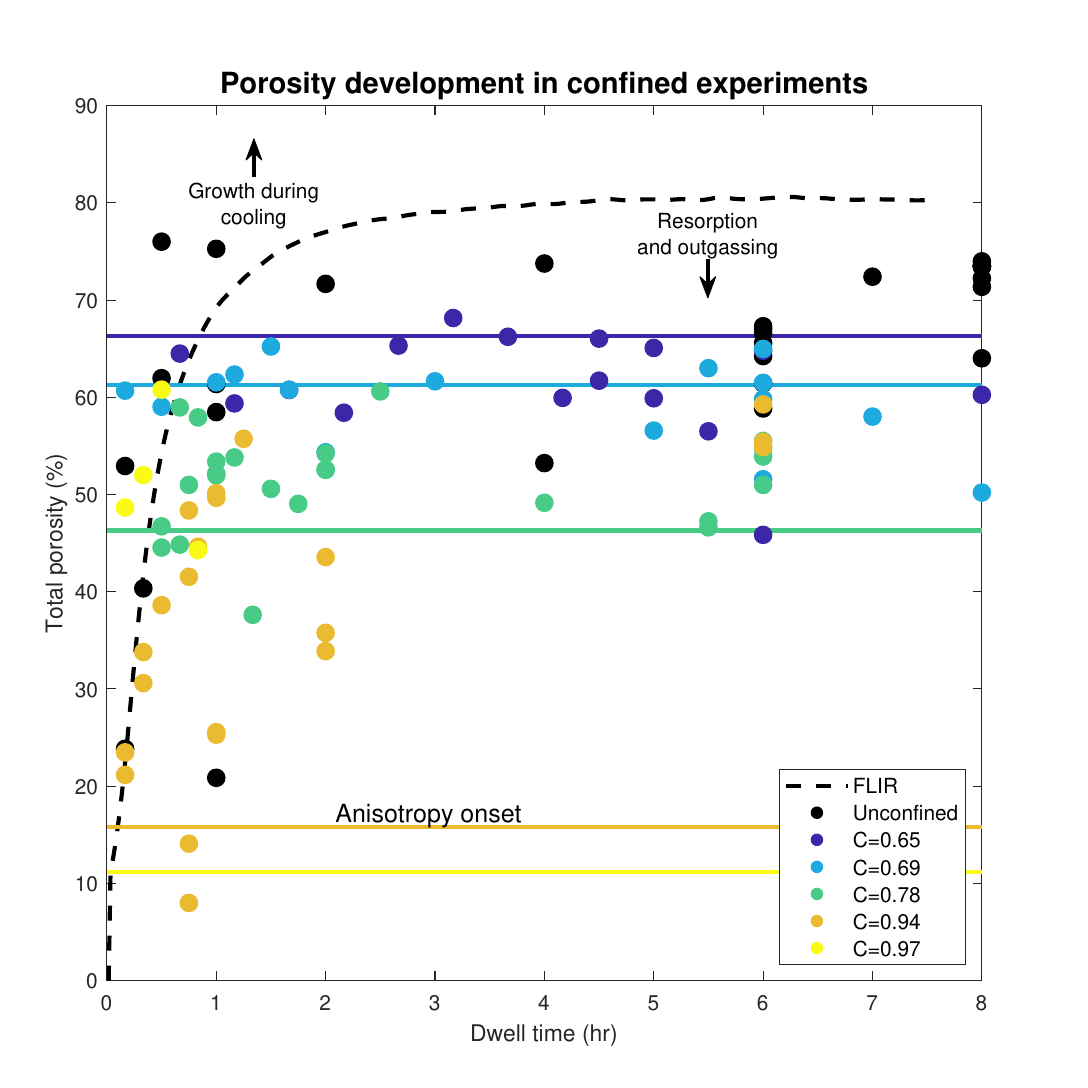}\\
\caption{Porosity measurements of post-experimental samples with variable dwell time and different degrees of confinement indicated by different colors. The average porosity measured from FLIR imaging for unconfined samples (black) is shown in the dashed line. From the processed FLIR results, we calculate the porosity required for the sample to reach the crucible resulting in anisotropy onset, shown in colored lines.}
\label{fig:anisotropic_results}
\end{center}
\end{figure}

Final total porosity of the confined samples is consistently lower than for unconfined samples, with larger differences for samples with more confinement (Fig. \ref{fig:anisotropic_results}). We expect the creation of porosity during initial vesiculation is similar between the confined and unconfined samples, except for samples that begin in close proximity to the crucible walls (C=0.97), which may experience increased heterogeneous nucleation at the contact region, resulting in faster initial vesiculation. The samples at long duration show reduced porosity as a result of minor gas escape.  \par 

%TC:ignore
%TC:endignore

\begin{figure}[]
\begin{center}
\includegraphics[width=5in]{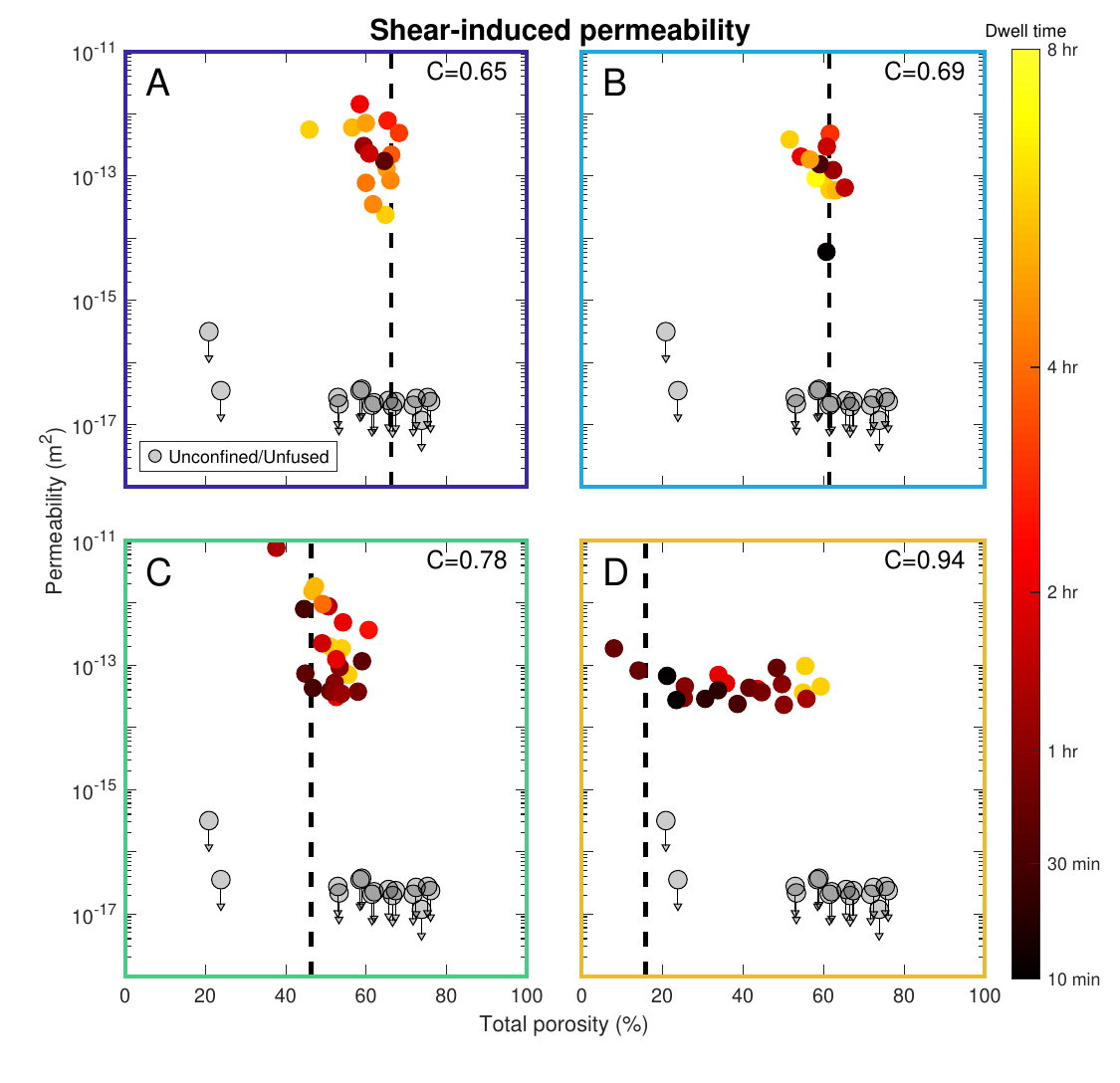}\\
\caption{Total porosity and permeability measurements for different degrees of confinement: A) C=0.65, B) C=0.69, C) C=0.78, and D) C=0.94 with colored frames consistent with the marker colors in Fig. \ref{fig:anisotropic_results}. Dashed lines indicate the calculated porosity of contact with the crucible. Measurements are colored by dwell time. Grey samples were unconfined or did not grow sufficiently to contact the crucible walls. Arrows indicate that measured values are maxima for permeability measurements terminated prior to equilibration.}
\label{fig:permeabilty_results}
\end{center}
\end{figure}

We observe two different regimes depending on the confinement ratio. In cases of low confinement (C=0.65 and 0.69), final total porosities rarely exceed the porosity required for the onset of anisotropy (61 and 66\%, respectively) and the majority of vesiculation occurs during the isotropic stage of growth (Fig. \ref{fig:permeabilty_results}). When the samples contact the crucible, porosity remains nearly constant afterward, while unconfined samples continued to grow after the same dwell times (up to $\sim$72\%). \par 

%TC:ignore
%TC:endignore

\begin{figure}[]
\begin{center}
\includegraphics[width=4in]{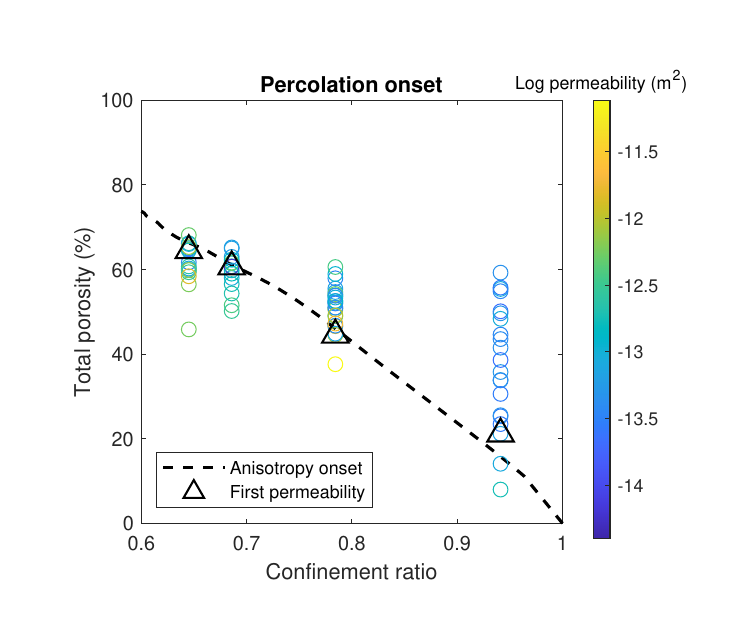}\\
\caption{Porosity resulting in significant permeability for different confinement ratios, colored by permeability. The black triangles highlight the experiments with the shortest dwell time that produced significant permeability ($>10^{-15}$ m$^2$). We interpret these samples to represent the onset of percolation. Samples at longer durations maintain permeability at lower porosity than the initial onset (Fig. \ref{fig:permeabilty_results}), which are visible below the black triangles in this figure. The dashed black line indicates the porosity required for the sample to first reach the crucible, but all samples shown here are coupled to the crucible.}
\label{fig:percolation_results}
\end{center}
\end{figure}

At higher confinement (C$\geq$0.78), the samples reach the crucible earlier and continue to increase in porosity after the onset of anisotropy. The onset of permeability is at lower porosity (46\%) with more confinement (Fig. \ref{fig:percolation_results}), but the permeability at that porosity is not necessarily higher. The samples at C=0.94 show a clear progression with increasing dwell time leading to higher porosity but permeability either decreases or remains stable with further vesiculation. The maximum porosity achieved in these experiments is lower than for the low confinement samples and the onset of permeability occurs at a lower porosity. \par 

In all cases, the lowest porosity sample which shows connectivity occurs after a longer dwell time and exhibits higher permeability than at which porosity is first established. We observe either no, or a mild inverse, correlation between porosity and permeability. This relationship is subtle, because porosity remains nearly constant between samples of the same confinement, especially in the isotropic-dominated regime. The measurements roughly overlap with the "explosive" region identified in \citet{Mueller2005}, and with existing compilations \citep[e.g.][]{Colombier2017,Lavallee2021}. \par 

%TC:ignore
%TC:endignore

\subsection{Bubble textures}
\begin{figure}[]
\begin{center}
\includegraphics[width=\textwidth]{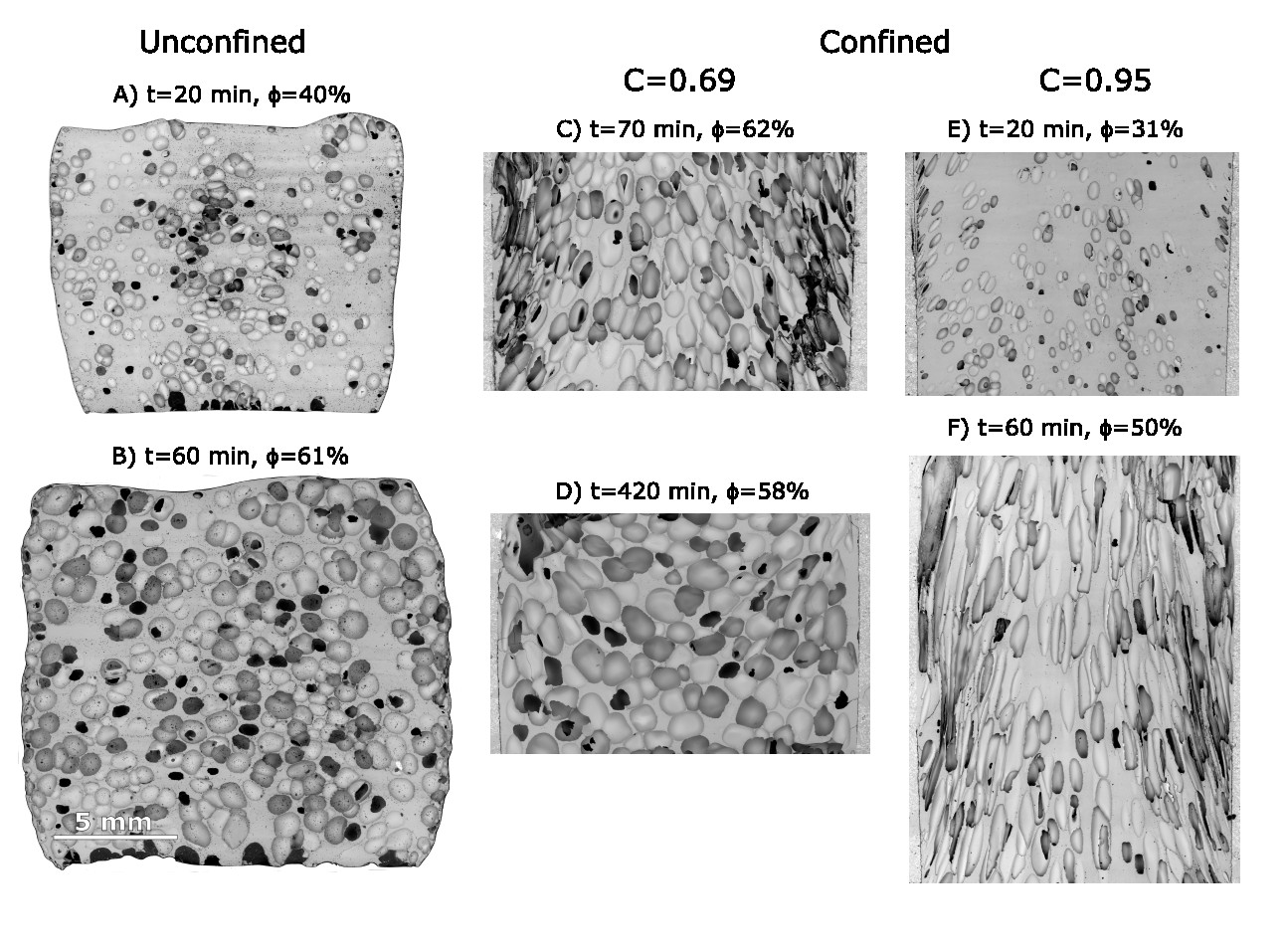}\\
\caption{Backscattered electron images of A-B) unconfined samples with increasing dwell time and total porosity and C-D) confined samples with at C) the beginning of shear, and D) after significant shape relaxation, E-F) samples with more confinement that show E) shear during early vesiculation, and F) volume loss after vesiculation during shear. Note the similar dwell times for B, C, \& F.}
\label{fig:SEM}
\end{center}
\end{figure}

BSE images of select cut samples (Fig. \ref{fig:SEM}) show the textural differences between bubbles in the unconfined and confined experiments. In the unconfined samples we see nearly equant bubble shapes with increasing bubble size at increasing dwell time, resulting in higher porosity. Between Fig. \ref{fig:SEM}B\&C, we compare two experiments with nearly the same dwell time and porosity, but with different confinement. The partially confined sample shows bubble elongation with highly distorted bubbles sub-parallel to the flow direction at the margins and with slightly elongate and larger bubbles in the center aligned with the flow direction. \par 

%TC:ignore
%TC:endignore

In contrast, for the sample that experienced much longer dwell time (420 min, Fig. \ref{fig:SEM}C), longer than the period of rapid vesiculation identified in the unconfined experiments of $\approx100$ min, shear has slowed or stopped enough for the bubbles to relax to near-round shapes again with slightly lower porosity. The bubble orientations suggest that the velocity in the center of the sample may have reversed direction with either gas loss or simply through shape relaxation of the interior bubbles. \par 

At higher degrees of confinement, the bubble elongation is present in even very small bubbles early in the vesiculation process. Fig. \ref{fig:SEM}F shows a similar dwell time again as panels B\&C, but shows higher aspect ratio bubbles, closer alignment to the flow direction, and lower total porosity suggestive of more efficient outgassing. \par 

\section{Discussion}
\subsection{Sources of uncertainty}
We observe a fair amount of scatter in our porosity measurements, especially for experiments at high confinement ratios. While we assume that the rind at the surface of the samples after the experiments isolates the interior porosity, fractures or breaking of the samples at the base plate may in some cases connect portions of the sample interior which could result in underestimates of the porosity, especially for samples with high interior connectivity. Additionally, the bulk porosity represents the entire sample, but we measure permeability in a horizontal slice after removing the upper and lower rind which may not be representative of the entire sample, which necessarily must have a vertical gradient in porosity and pore geometry. Indeed the connected porosity on some cut samples exceeds the bulk porosity of those samples prior to cutting, which could reflect either connectivity in the bulk porosity measurement or that the central slice avoids a low porosity region at the sample ends as a result of inhibited growth or diffusive volatile loss. Finally, bulk porosity is measured using changes between pre-experimental and post-experimental volumes, while the connected porosity determination is conducted on a comparison between post-experimental cut volumes and a calculation of the cylindrical volume which has potential error from sample faces that are not parallel (typical variation in height around the sample is 0.05 mm leading to an uncertainty in volume and porosity of approximately 0.2-1\%), and for the unfused samples variation in radius (of order 1 mm) which introduces an additional error (typically an overestimate of connected porosity) of approximately 1-3\%. \par

Minor disagreements between the final measured samples and syn-experimental FLIR observations suggests that we may be missing continued deformation during the early stages of cooling which include bubble shrinkage and a decrease in melt volume which could modify the bubble textures, and allow for fracturing or transient disruption of the dehydrated rind which could allow for minor outgassing during cooling. We observe an approximately 5-10\% difference in final measured porosity and the FLIR measurements. Thermal contraction of the melt phase from the experimental temperature of 1009 $^\circ$C to room temperature (20 $^\circ$C) typically results in a volume reduction of approximately 1\%. Similarly, the volume of the gas phase will decrease during cooling. If we consider only the interval in which melt is relaxed from 1009 $^\circ$C to 900 $^\circ$C, we would expect an approximately 10\% change in water vapor volume usign the ideal gas law, suggesting that final estimates of porosity may be a slight underestimate for the porosity at high temperatures. This also helps explain (along with changes in aspect ratio and outgassing) why some samples are coupled to the crucible walls, despite having porosities slightly below the predicted porosity required to touch the crucible walls. \par

\subsection{Permeability hysteresis}
To a first order, permeability is expected to increase with increasing porosity, with shear decreasing the threshold for connectivity. We clearly observe the second prediction. So why, then, do we observe an apparent decrease in permeability with increasing total porosity at a given confinement ratio (e. g., Fig. \ref{fig:permeabilty_results}D)? The Kozeny-Carman model idealizes permeable flow through porous media as flow confined to pore channels that transverse the sample through potentially complicated paths that offer resistance proportional to the friction along the channel walls, offering a direct link between pore geometry and permeability. Many approximations exist for various geometries, for example: 
\begin{equation}
    k = K\frac{\phi^{m}}{(1-\phi)} \: , 
\end{equation}
from \citet{Costa2006}. In practice, calculations for $K$ and $m$ only exist for highly simplified geometries and are fit empirically to pass through data. We expect these to vary with the connected porosity, characteristic bubble and pore sizes (conductance), tortuosity, and specific surface area (itself a function of the preceding). Tortuosity is typically defined according to the ratio of the path length the percolating fluid must travel compared to the sample length, where increasing tortuosity acts as an additional impediment to flow and decreases permeability. \par

\begin{figure}[]
\begin{center}
\includegraphics[width=4in]{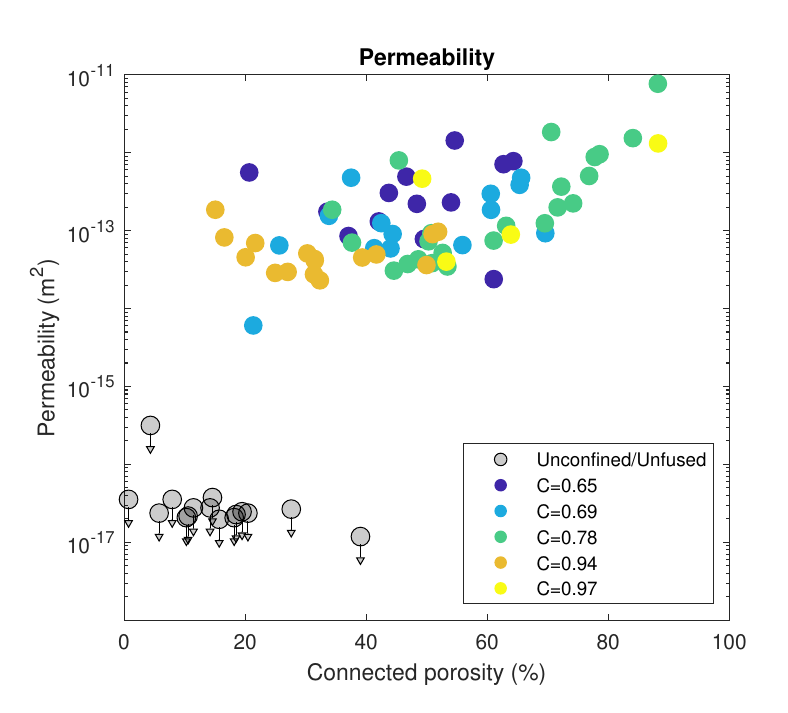}\\
\caption{Permeability as a function of connected porosity and confinement ratio (color). Arrows indicate that measured values are maxima for permeability measurements terminated prior to equilibration. We observe a distinct separation between sheared and unsheared (unconfined/unfused) samples.}
\label{fig:connectivity}
\end{center}
\end{figure}

The compilation of connectivity data in \citet{Colombier2017} shows that in natural samples, especially from lava flows which experience more shear, connectivity typically rapidly increases over a narrow interval and then remains nearly constant between about 0.8-1 (vol/vol). That is, above the threshold for connectivity, pores remain nearly all connected with a hysteresis in which connectivity persists during bubble relaxation or collapse, even to low porosities. Indeed, our sheared samples show nearly complete connectivity, while the unsheared samples uniformly have low connectivity ($\lesssim$0.5). When we compare connected porosity with permeability, we do observe a broad positive correlation across all degrees of confinement in sheared samples (Fig. \ref{fig:connectivity}), with a clear separation from the unsheared samples. However, within the C=0.94 samples, the pores are nearly all connected, suggesting that the mild inverse correlation between total porosity and permeability is controlled by a factor other than connectivity. \par 

We may expect that tortuosity, in the direction of the fabric, should decrease with the increasing anisotropy of the sample \citep{Wright2009a}. While we discuss permeability as a single value for each sample in this paper, in reality permeability is a tensor and depends on the direction of flow. Because all our samples are measured in the same orientation, with flow through the cylindrical crucible, the direction of permeability measurement is parallel or sub-parallel to the direction of bubble elongation. We might therefore expect that with increasing anisotropy (shear), samples should display higher permeability even at the same porosity and connectivity. This would suggest that samples with the same porosity but greater confinement should show higher permeability. This trend is not evident in the data. Furthermore, given the continued high permeability at the longest times which have experienced significant bubble shape relaxation (Fig. \ref{fig:SEM}D), tortuosity does not seem to play a major role in explaining the variability in permeability of these samples, perhaps because the pore connectivity structure partially remains during relaxation without dramatically increasing tortuosity. Other studies have found a low correlation between tortuosity and permeability \citep[e.g.][]{RezaeiNiya2018}. \par 

%TC:ignore
%TC:endignore

Finally, we turn to conductance: the cross-sectional area of the channels through which fluid flows. In bubbly suspensions, this usually relates to the aperture size between connected bubbles, which increases with increasing bubble size and coalescence \citep{Saar1999}. We do observe a continued increase in bubble size with time in our experiments. Additionally, highly elongate bubbles will potentially have smaller pore apertures compared to a rounded bubble of the same volume/equivalent diameter. This provides a mechanism for maintaining high permeability in the low confinement ratio experiments and explains some of the hysteresis in which shear promotes the initiation of widespread connectivity which is maintained during bubble shape relaxation at longer dwell times. This is also supported by the higher permeability observed at onset in the low confinement samples which must already have large bubbles at the time of shear-induced connectivity. \par 

Changes in the pore geometry would result in different apparent fitting factors in the Kozeny-Carman relationship for each sample with its own unique history. We would not expect one curve to explain the variability in our data, but rather a family of curves specific to the heating and shear history experienced by different samples. For example, permeability through spherical pores should scale with the aperture radius to the fourth power \citep{Saar1999}, meaning our measured variability in the permeability of $\sim$1-2 orders of magnitude at a given porosity could be explained by a factor of only $\sim$2 increase in aperture radius. Additionally, surface area for a pore increases with increasing elongation, which at the same tortuosity would approximately halve permeability for bubbles with an aspect ratio of 10 compared to spheres. Further investigation of the relationship between pore elongation and pore throat size and resultant permeability appears a fruitful direction for future work in volcanic materials. \par

\subsection{Regimes of bubble growth and outgassing}
Like many other studies on permeability development in magmas and lavas, we are limited to measurements of permeability in cooled and solidified rocks which necessarily do not have the same dynamics as a melt. Each sample represents a snapshot of a particular time and pore geometry. Fortunately, we can combine our permeability measurements with the development of porosity to understand gas transport at high temperatures. If the melt is relaxed (as suggested by the observed changes in bubble shape), then the volume change of the pore space reflects the mass balance of water vapor in the bubbles. Diffusion of water from the melt is the only source of water mass for bubble growth in these experiments, and should be relatively similar across different degrees of confinement, with perhaps slightly faster diffusion into elongate bubbles which have a higher surface area. \par 

We can consider the relative contribution of different water loss mechanisms. The experiments of \citet{vonAulock2017} demonstrate the importance of the thin dehydrated rind in retaining vapor bubbles in the melt phase. Water diffusion through this rind can reduce the porosity of the sample with sufficient time, and particularly for small samples \citep{Weaver2022}. If dehydration through this rind was the dominant water loss mechanism, we would expect that the samples with the greatest surface area exposed to the air (unconfined samples) would show the greatest gas loss, this volume loss would be apparent in Fig. \ref{fig:isotropic_results}, and the samples should show decreased porosity preferentially at the sample edges in Fig. \ref{fig:SEM}. Since this is not the case, the gas must be lost via permeable pathways connected to the exterior of the sample. Perhaps shear along the conduit margins disrupts the rind formation and allows for some permeable outgassing. \par 

\begin{figure}[]
\begin{center}
\includegraphics[width=3in]{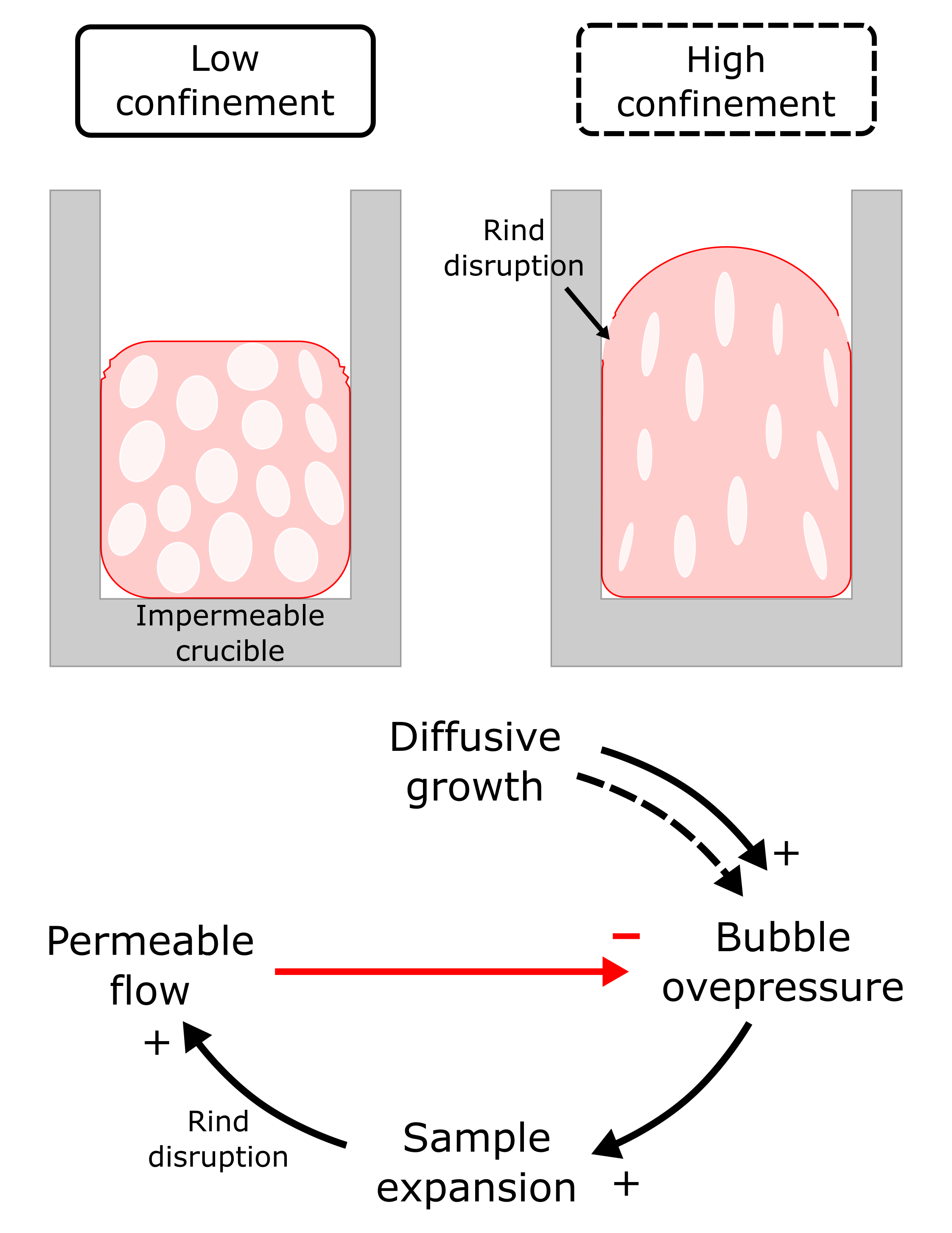}\\
\caption{Feedbacks between bubble growth due to water diffusion and outgassing through permeable flow along the disrupted rind. Black arrows represent reinforcing relationships and red arrows represent mitigating relationships, the dashed arrow highlights the extra role of diffusive growth in high confinement geometries. }
\label{fig:feedbacks}
\end{center}
\end{figure}

The gas loss in our partially confined experiments suggests that permeability and connectivity developed rapidly after the onset of shear. In the low confinement experiments, this permeability development during the end stages of vesiculation seems to be sufficient to relieve bubble overpressure and halt further growth. If rind disruption is a crucial part of connecting the sample interior to the surrounding atmosphere, we might expect permeable gas flow to end when the sample expansion stops and the rind is able to reform and persist. For these samples, system-scale permeability is likely a transient behavior at high temperatures with a feedback between bubble growth and permeable gas loss which are roughly balanced (Fig. \ref{fig:feedbacks}). As the samples continue to mature with bubble coalesce and shape relaxation, which changes the permeability of the cooled and cut samples, they are likely isolated from the atmosphere by the rind at long times. \par

%TC:ignore
%TC:endignore

When the onset of shear occurs early in the vesiculation, bubble growth continues after shear, albeit at a suppressed rate, in which water diffusion into the bubble is still faster than water loss. This allows continued flow of the material, with continued deformation potentially reducing the sealing effect of the rind on the sample's free surface. These high confinement samples show a slower vesiculation path and reach lower final values of porosity because they maintain a partially open system for a longer portion of their history. \par 

Interestingly, the fortuitous choice of basalt crucibles may play an important role in retaining the bubbles except through the torn rind at the sample surface. The basalt itself has a very low permeability which limits outgassing along the bottom and sides of the sample. A porous ceramic crucible may have allowed such gas escape \citep{Seropian2022}. This highlights the importance of understanding the rock properties of conduit walls in natural systems which may play an important mediating role in allowing gas escape during magma ascent, which has been highlighted by numerous conduit ascent models \citep[][e. g.]{Jaupart1991,Woods1994}. \par 

\section{Conclusions}
Our experiments document the importance of shear on establishing bubble connectivity and permeability in silicate melts. We find that during vesiculation in the absence of shear, the percolation threshold is very high, in fact all of our samples, even up to our maximum porosity of 76\%, show permeability less than $10^{-16}-10^{-15}$. After only small amounts of shear, connectivity is established quickly. In samples that reached 60-70\% vesicularity, permeability is established upon the onset of shear and has overall high values of permeability between 2$\times$10$^{-14}$ and 2$\times$10$^{-12}$. In samples with lower porosity and more rapid shear, permeability is established at lower porosity ($<$20\%), but with a lower permeability 2$\times$10$^{-14}$-2$\times$10$^{-13}$. Under shear, bubble connectivity is established and remains present even after the end of shear and significant bubble shape relaxation. Instead, permeability in sheared samples is likely a function of both connected porosity and conductance with changing bubble size and aspect ratio. \par

\section{Data availability}
Experimental data will be uploaded to Zenodo upon acceptance. 

\section{Author contributions}
JB: Investigation, writing - original draft, writing - review \& editing, visualization; JS: Investigation, writing - preliminary draft; JW: Investigation; JEK: Conceptualization, methodology, writing - review \& editing, supervision, funding acquisition; AL: Conceptualization, methodology, writing - review \& editing, supervision; YL: Conceptualization, methodology, writing - review \& editing, supervision, funding acquisition.

\section{Funding sources}
This study was financially supported by a Consolidator grant from the European Research Council (ERC) on Magma Outgassing During Eruptions and Geothermal Exploration (MODERATE No.101001065), a Natural Environment Research Council (NERC) standard grant (NE/T007796/1) and NERC EAO Doctoral Training Partnership (NE/L002469/1). 

%TC:ignore
\newpage
\bibliography{references}
%TC:endignore

\end{document}

%% file: Table_composition.tex
\begin{table}[h!]
\fontsize{10}{12}\selectfont
\centering
\begin{tabular}{|l|r|}
    \hline
    Oxide & wt \% \\
    \hline
    % coordinate variables
    SiO$_2$ & 48.40 \\
    TiO$_2$ & 3.70 \\
    Al$_2$O$_3$ & 14.35 \\
    Fe$_2$O$_{3,\text{tot}}$ & 14.60 \\
    MnO & 0.22 \\
    MgO & 4.58 \\
    CaO & 9.13 \\
    Na$_2$O & 3.81 \\
    K$_2$O & 1.00 \\
    P$_2$O$_5$ & 0.59 \\
    SO$_3$ & 0.01 \\
    LOI & -0.69 \\
    total & 100.39 \\
    \hline
    H$_2$O & 0.12 \\
    \hline
\end{tabular}
\caption{Major element composition from XRF analysis and water concentration from FTIR.}
\label{tab:composition}
\end{table}